\documentclass[conference]{IEEEtran}

\ifCLASSINFOpdf
\else
\fi
\usepackage[hyphens]{url}
\usepackage{hyperref}
\begin{document}

\title{User Negotiations of Authenticity, Ownership, and Governance on AI-Generated Video Platforms: Evidence from Sora}

\author{
\IEEEauthorblockN{
Bohui Shen$^{1}$,
Shrikar Bhatta$^{2}$,
Alex Ireebanije$^{3}$,
Zexuan Liu$^{4}$, \\
Abhinav Choudhry$^{4}$,
Ece Gumusel$^{4}$,
Kyrie Zhixuan Zhou$^{5}$}

\IEEEauthorblockA{$^{1}$Independent Researcher \quad $^{2}$Stevenson High School \quad $^{3}$MTN Nigeria\\
$^{4}$University of Illinois Urbana-Champaign \quad $^{5}$University of Texas at San Antonio}

\IEEEauthorblockA{
zosh.olivia@gmail.com, shrikarbhatta@gmail.com, ireebanijealex@gmail.com,\\
zexuanl5@illinois.edu, ac62@illinois.edu, eceg@illinois.edu, kyrie.zhou@utsa.edu}
}




\maketitle

\begin{abstract}
As AI-generated video platforms rapidly advance, ethical challenges such as copyright infringement emerge. 
This study examines how users make sense of AI-generated videos on OpenAI’s Sora by conducting a qualitative content analysis of user comments. 
Through a thematic analysis, we identified four dynamics that characterize how users negotiate authenticity, authorship, and platform governance on Sora. 
First, users acted as critical evaluators of realism, assessing micro-details such as lighting, shadows, fluid motion, and physics to judge whether AI-generated scenes could plausibly exist. 
Second, users increasingly shifted from passive viewers to active creators, expressing curiosity about prompts, techniques, and creative processes. 
Text prompts were perceived as intellectual property, generating concerns about plagiarism and remixing norms. 
Third, users reported blurred boundaries between real and synthetic media, worried about misinformation, and even questioned the authenticity of other commenters, suspecting bot-generated engagement. 
Fourth, users contested platform governance: some perceived moderation as inconsistent or opaque, while others shared tactics for evading prompt censorship through misspellings, alternative phrasing, emojis, or other languages. 
Despite this, many users also enforced ethical norms by discouraging the misuse of real people’s images or disrespectful content. 
Together, these patterns highlighted how AI-mediated platforms complicate notions of reality, creativity, and rule-making in emerging digital ecosystems.
Based on the findings, we discuss governance challenges in Sora and how user negotiations inform future platform governance.
\end{abstract}


\IEEEpeerreviewmaketitle

\section{Introduction}

Recent advances in generative artificial intelligence (AI), particularly in text-to-video systems, have intensified debates over authenticity, ownership, and platform governance across social media. 
Platforms such as OpenAI’s Sora\footnote{\url{https://sora.chatgpt.com/}} mark a new phase in synthetic media production, enabling the generation of photorealistic video from natural language prompts. 
Users can simulate realistic worlds, manipulate physical dynamics, and produce creative content without traditional filmmaking expertise.
As these systems blur the boundary between the real and the artificial, scholars have increasingly examined how users evaluate authenticity \cite{chen2025examining, farooq2025deciphering, riter2025representation} and negotiate authorship \cite{wang2025research, khadka2025navigating, wang2025excavating}.

While existing research provides substantial insight into authenticity judgments, ownership ambiguity, and governance challenges in generative AI environments, there is limited empirical work that captures real-time user negotiations of these issues specifically within AI video platforms. 
As a result, legal and psychological perspectives on AI-generated content (AIGC) often remain disconnected from analyses of everyday user practice. 

Although Sora remains a relatively recent platform, teaser releases and media demonstrations have generated substantial public discourse on its capabilities and risks \cite{liu2024sora, mogavi2024sora}. 
Early commentaries on Sora emphasize its unprecedented realism, creative potential, and disruptive implications for filmmaking, advertising, and social media production \cite{zhou2024sora}. 
However, scholarly work on public perceptions of Sora is still emerging and often indirect, drawing on broader studies of generative AI trust, acceptance, and ethical evaluation rather than systematically analyzing user-generated data.
Such gaps suggest the need for qualitative, discourse-driven research that examines how users articulate authenticity, authorship, and governance in relation to specific platforms such as Sora.

The present study addresses these gaps by qualitatively analyzing user comments on Sora videos to understand how authenticity, authorship, and governance are negotiated in everyday discourse.
In particular, we answered the following two research questions:
\begin{itemize}
    \item How do users act as content consumers and creators on AI-generated video platforms such as Sora?
    \item How do users cope with ethical and governance challenges on AI-generated video platforms such as Sora?
\end{itemize}

Our analysis identifies four patterns in how users engage with AI-generated videos. 
First, users acted as critical evaluators of realism, inspecting micro-details, such as lighting, shadows, fluid motion, and physics, to judge plausibility. 
Second, they increasingly shifted from passive viewers to active creators, expressing interest in prompts and techniques while debating the ownership and remixing of text prompts as intellectual property. 
Third, users navigated blurred boundaries between real and synthetic media, expressing concerns about misinformation and even questioning whether fellow commenters were bots.
Fourth, they contested platform governance: some viewed moderation as inconsistent or opaque, while others shared strategies for circumventing prompt filters. 
At the same time, users collectively enforced ethical norms by discouraging harmful uses of real individuals’ likenesses.

This paper makes three major contributions.
First, we offer one of the earliest empirical analyses of real-time user negotiations on AI-generated video platforms by qualitatively examining everyday discourse surrounding OpenAI’s Sora. 
Second, we reveal how Sora users collectively construct emerging norms around authenticity, authorship, and ownership, ranging from physics-based realism assessment to prompt protection, plagiarism accusations, and remix ethics. 
These findings advance theoretical understanding of psychological ownership and authorship in generative video ecosystems.
Third, we uncover a dual pattern of governance practices: users simultaneously circumvent formal moderation through linguistic evasion while enforcing informal ethical norms within the community. 

\section{Related Work}

We discuss several lines of research, including (1) AI-generated content on social media and perceived authenticity and (2) debates of ownership and authorship of AI-generated media.


\subsection{AI-Generated Content on Social Media and Perceived Authenticity}

AI, particularly generative AI, brings both benefits (e.g., rapid content creation, personalized recommendations, harmful content filtering) and risks (e.g., misinformation/disinformation, social bots, covert manipulation) to social media \cite{pan2025complexity}.
AI personalizes social media marketing strategies and impacts customer experience, enabling tailored content delivery and improved engagement \cite{beyari2025role}.
Kubovics finds that acceptance of AI-generated marketing content is conditioned by perceived usefulness, ethical transparency, and user trust in the underlying technology \cite{kubovics2025influencing}. 
Other business applications of AIGC include virtual influencers and cloning personas \cite{su2025introduction}. 
Social media algorithms are found to amplify AI posts and create engagement cycles that boost viral AI-generated content \cite{bond2024ai}.

AIGC has been found to have a profound impact on social media users.
AI-generated images in social media advertising affect user engagement in digital marketing \cite{sharma2024user}.
A comprehensive study of Pixiv \cite{wei2024understanding}, an online community for artists sharing illustrations, examines how AI-generated content affects artist communities and creative feedback dynamics.

A central concern in the literature on AI-generated media is how users assess the authenticity of synthetic content in environments already saturated with manipulated visuals and misinformation \cite{drolsbach2025characterizing}. 
Park et al. investigate user perceptions of AI-generated versus human-created content on Instagram, highlighting challenges in distinguishing between AI- and human-generated content and concerns about misinformation, privacy breaches, and declining trust in digital media \cite{park2024ai}.
An experimental study with 680 U.S. participants shows that AI tools increase user engagement and content volume but decrease perceived quality and authenticity \cite{moller2025impact}.
Farooq and de Vreese show that exposure to AI-generated disinformation images complicates authenticity judgments, particularly when detection tools are inconsistently applied \cite{farooq2025deciphering}. 
In contrast, Ilie shows that in academic contexts, users apply authenticity criteria not only to outputs but to the process through which content is generated, indicating that public evaluations are increasingly process-sensitive rather than outcome-focused \cite{oana2025ethics}.
In archival and information contexts, Riter et al. argue that AI-generated and AI-curated media destabilize traditional notions of representation and authenticity, especially when synthetic materials circulate without clear provenance \cite{riter2025representation}. 
Their work emphasizes that authenticity is increasingly co-produced by platforms, algorithms, and user communities who collectively determine what is perceived as ``real.''

On the other hand, traditional notions of authenticity, too, have previously faced challenges arising from altered media. 
Manipulation in photography was even used by Mussolini, and there have been varying degrees of receptiveness to the discovery of photographic alteration, ranging from firings and resignations to more benign reactions \cite{sharma2017analysis}. 
Twenty years ago, after Adobe Photoshop had become mainstream, Emme and Kirova discussed the advantages of digital alteration while also acknowledging the risks: ``Abstraction through manipulation of the subject, setting, or materials of production results in an image that can serve as an ideal type. Advertising photography is a ubiquitous, if somewhat problematic example of this'' \cite{emme2005photoshop}.
There is evidence, though, that people have been adapting their visual presentation to become more generic, which was termed ``cyborgian face" or ``Instagram face" by Tolentino \cite{tolentino2019age}.
Many people have even begun trusting artificially generated recommendations more in specific contexts, such as AI in the context of material goods \cite{jin2025artificial} and virtual influencers in the context of digital products \cite{choudhry2022felt}. 
The general loss of faith in online social media discourse was underscored in a 2023 Gartner survey that found that half of respondents wanted to limit their social media use in the near future due to misinformation, toxic user bases, and bots \cite{gartner2023socialmedia}. 
However, AI-generated content evoked even stronger negative sentiment, with 70\% feeling that it would harm user experience on social media \cite{gartner2023socialmedia}.  

Labeling is an effective intervention to help people distinguish AI-generated content.
Chen et al. demonstrate that labeling practices and the perceived stakes of content significantly shape user judgments of AI-generated images on social media \cite{chen2025examining}. 
Li and Yang \cite{li2024impact} found that AIGC labels help distinguish AI content but have minimal impact on user perceptions.

\subsection{Content Ownership: Intellectual Property Rights in AI-Generated Media}

Generative AI disrupts conventional models of authorship by introducing non-human agents into the creative process. 
Wang argues that copyright law struggles to accommodate AI-generated works due to the absence of clear human authorship, leading to regulatory ambiguity over ownership, liability, and economic rights \cite{wang2025research}. 
Khadka similarly emphasizes that accountability frameworks remain underdeveloped for AI-created content, creating legal grey zones around attribution and misuse \cite{khadka2025navigating}. 
The manner of data gathering and the purpose, whether it is commercial or otherwise, emerge as key points in current legal battles over copyright \cite{ismantara2025lawfulness}. 
Text-to-video, including Sora, has been highlighted for copyright infringement and ownership concerns already in the literature \cite{chunguang2024investigation}.   

Beyond formal legal ownership, psychological ownership has emerged as a critical dimension of user engagement with digital platforms. 
Wang et al. show that users develop a sense of psychological ownership over digital content through participation, customization, and social recognition \cite{wang2025excavating}. 
In the context of AI-generated video, this translates into users perceiving text prompts, editing choices, and creative direction as personal intellectual contributions. 
As a result, prompt plagiarism, unauthorized remixing, and content scraping are frequently contested within user communities, even in the absence of formal legal protections. Providing intellectual property rights to prompts is now being proposed by some authors \cite{van2023protect} but this is still a very young space. 

\section{Methodology}

\subsection{Data Collection}

We collected user discussions from the Sora platform to examine how users perceive and interact with AI-generated videos. 
To ensure sufficient conversational depth, we included only videos with 10 or more user comments, using a heuristic agreed upon by the research team. 
Following this criterion, we identified 41 videos, yielding 254 comments, which enabled us to reach theoretical saturation in the analysis.
Note that we ignored overly short comments during data collection because they were not deemed helpful for our later analysis. 
We did not seek IRB approval because all data, including Sora videos and comments, are publicly available.

Data collection and analysis were conducted iteratively and concurrently. 
As new videos and comments were added to the dataset, preliminary coding was performed in parallel, allowing early patterns to inform subsequent sampling decisions. 
This iterative approach enabled us to refine emerging themes and focus on comment threads that revealed rich user practices or ethical concerns.

We continued this iterative process until theoretical saturation was reached, i.e., when additional data no longer yielded new themes or meaningful conceptual insights.
At this point, the dataset was deemed sufficient to support a stable thematic structure and to answer the research questions guiding the study.

\subsection{Data Analysis}

We conducted a collaborative thematic analysis \cite{braun2006using}, involving two research team members, to identify how users evaluate, interpret, and negotiate AI-generated videos on Sora. 
Analysis began alongside data collection: as new comments were added to the dataset, we performed open coding to capture preliminary concepts related to realism assessment, creative practices, platform governance, and ethical concerns. 

During the analysis, we examined how individual comments related to one another across the corpus, enabling us to trace broader patterns in user perception and behavior. 
Throughout the process, we wrote analytical memos to capture emerging themes and subthemes.

Regular meetings were held across the entire research team to discuss the themes identified by the two coders and reach consensus.
In the end, we identified four major themes: users as critical content consumers, identity transformation from viewers to creators, blurred reality, and contesting platform governance.
In the results section, we will use raw comments, with emojis removed, to illustrate our findings.

\section{Results}

In this section, we report results on how users serve as critical content consumers and active content creators, perceive the blurring of reality, and contest platform governance. 
Negotiation is a central theme in our results, with users negotiating the authenticity of content, ownership of prompts, and platform governance in the human-algorithmic ecosystem.

\subsection{Users as Critical Content Consumers}

\subsubsection{Positive Appraisal of Video Quality and Creativity}

A number of user comments appraised the quality and realism of the videos positively: \textit{``Simple yet genius.''} 
\textit{``That was truly unexpected, well done''} 
\textit{``Brilliantly executed.''} 
\textit{``Bro it’s so crazy how real Ts looks''} 
\textit{``Now you got some of the best remixes I’ve ever seen and DEFINITELY the best like of remixes in general fr FRRR''}

Given the level of reality in the videos, people were actively discussing the content of the videos, treating them as real videos: 
\textit{``It actually explodes. Sometimes it doesn’t do it until you go to crack it.  But it still explodes.''}

Realism often provoked immediate physical reactions. 
Users described being startled or disoriented, revealing that AI-generated visuals can elicit embodied responses similar to those triggered by real stimuli.
\textit{``That actually jumpscared me.''}
\textit{``STOP THAT SCARED ME.''}
\textit{``Heart attack moment.''}
\textit{``I literally flinched when it exploded.''}
\textit{``It felt so real I looked away for a sec.''} 

Creativity of the videos was highlighted: 
\textit{``Lmfao very creative.''}
Some attributed creativity to the prompters/creators instead of AI: 
\textit{``Fr I want to play with him and WHY WHY WHYYY CANT I COME UP WITH THESE IDEAS TOO''}

\subsubsection{Verification of Realism through Physical Consistency}

Users on Sora do not passively accept the realism of the videos on the platform. 
Instead, they actively examine how physical laws operate in each scene to determine whether what they see could plausibly exist in the real world. 
When delineating realism of the videos, they focus on micro-details, such as light, motion, fluid dynamics, and collision physics, rather than overall aesthetics. 
When these physical laws appear coherent, users describe the video as ``real'' or ``convincing.''
\textit{``The water physics on this video are amazing.''} 
\textit{``When the beer bottle drops it creates a splash on a moving wave.''}
\textit{``Even the shadow moves right when the camera pans.''} 
\textit{``The reflection on the floor is too real to be AI.''} 
\textit{``That bottle doesn’t explode until you crack it -- just like real life.''} 
Other users take an opposite approach by pointing out unrealistic details, serving as critical viewers rather than admirers:  
\textit{``This is so fake. Ducks don’t have hands.''} 
\textit{``I like how the jump cut makes this look faked.''} 
This duality between admiration and scrutiny shows that users are not passive consumers. 
Instead, they participate as ``amateur critics,'' collaboratively identifying what makes AI-generated realism convincing or flawed. 
Such analytical engagement transforms viewing into a participatory and interpretive practice.
Further, through such micro-scrutiny, users demonstrate an emerging form of physics literacy, i.e., an intuitive ability to verify realism by observing how matter behaves in AI-generated videos.

\subsection{Identity Transformation -- From Viewers to Creators}

Sora users tend to shift from passive viewers to active creators in our analysis. 

\subsubsection{Aspiring to Become Creative Creators}

Many comments reveal users’ curiosity about how to generate similar videos, such as:  
\textit{``Nice to meet you, that’s a very nice video. Please tell me the prompt.''}  
\textit{``OK, but what did you type to make this video? That’s what I wanna know!!!''}  
\textit{``How do you do the stitching thing?''} 
\textit{``How do I also add my own voice to a character’s lip movement?''} 
Some people specifically inquired about the ways to generate videos based on text prompts: 
\textit{``Is this the original prompt for this vid''}
\textit{``guys i tried it and it doesent work, can somebody help me?''}
These comments demonstrate a strong desire to learn, imitate, and master the skills behind generative AI content creation, such as prompting.

Some users further reflect on people's creative capabilities, expressing admiration for other creators and self-doubt:  
\textit{``Why can’t I think of this kind of stuff.''} 
This illustrates a broader shift in identity -- from mere audience members to aspiring content producers -- as users internalize generative creativity as part of their digital participation.

\subsubsection{Negotiating Authorship with AI -- The Value of Prompts}

As users gain technical fluency, the focus of discussion shifts from what the AI produces to who owns the act and products of creation. 
The Sora community reveals an emerging struggle between the platform’s open remix culture and individuals’ growing sense of authorship.

Prompts are seen as creative capital, i.e., users perceive textual prompts as the central contributor of creative labor. 
In their views, the right combination of words determines the success of a video, making the prompt itself a form of intellectual property: 
\textit{``Protect your prompt, people will steal it.''}
\textit{``Don’t post it or they’ll copy.''}
\textit{``You should sell that prompt.''}
These comments show that prompts have become a symbolic marker of expertise and ownership, and that users therefore view them as creative craftsmanship rather than mere instruction.

While Sora encourages users to ``remix'' existing videos, the community often interprets this act as a violation of authorship norms. 
Users accuse others of copying prompts, reproducing ideas, or exploiting the original creator’s effort:
\textit{``That’s my exact prompt!''}
\textit{``Wow thanks for stealing my content.''}
\textit{``You just copied my Russian channel.''}
\textit{``Tag me if you remix, don’t pretend it’s yours.''}
\textit{``Please credit the original creator.''}
These comments reveal an ongoing negotiation between platform affordances and prompt ownership. 
While remixing is technologically encouraged, users impose informal social norms that distinguish between learning from and stealing from others.

\subsection{Blurred Reality -- From Videos to Users}

\subsubsection{Indistinguishable Genetative Videos}

Many users struggle to distinguish AI-generated content from real footage. For example, commenters write: 
\textit{``I am believing this is real, incredible.''}  
\textit{``Genuinely forgot I was watching Sora and thought this was real.''} 
\textit{``Some say this is AI but this real life.''}  
\textit{``This genuinely looks like something that would happen irl.''}
\textit{``Forgot this was AI for a second.''}

Some users even describe moments of emotional immersion despite realizing later that the videos were synthetic:  
\textit{``Dude we’re on Sora and I forgot there’s only AI and I’ve been watching it thinking about the people in the house and what they’re gonna think. Then I realized it was AI.''} 
Such comments reveal the perceptual ambiguity and affective entanglement caused by AI media. 
Users’ sense of ``reality'' oscillates between authenticity and simulation, demonstrating how generative videos can challenge traditional boundaries of perception and belief.

\subsubsection{Concerns of Misinformation and Beyond}

Users extend their concern outward, speculating how less media-literate audiences might misinterpret AI-generated content. 
They explicitly mention Facebook users and older adults as examples of vulnerable groups who might be deceived:
\textit{``This is definitely fooling some Facebook moms.''} 
\textit{``Post this on Facebook and 80\% would believe it.''} 
\textit{``Someone removes the watermark and people will think it’s real.''}
\textit{``This is gonna do numbers on Facebook.''} 
\textit{``I’m sobbing some old lady is gonna share this.''}
\textit{``My mom would totally fall for this.''}
As seen, users assess not only their own perceptions but also the public’s potential vulnerability, positioning themselves as informed witnesses in a confused media landscape.

\subsubsection{Skepticism of Community Authenticity}

Users' uncertainty extends from videos to people. 
Many question whether the comments they read, or even the users they interact with, are real. 
This collective doubt signals a deeper crisis of trust within AI-mediated platforms.

Users frequently express disbelief that other commenters are real. 
The boundary between human and automated participation becomes blurred, producing an uncanny sense of social emptiness:
\textit{``Some of these comments make me think the people who use this app are also AI.''}
\textit{``I’m convinced all of these comments are bots.''}
\textit{``Half these profiles look generated.''}
\textit{``We’re in a loop of AIs commenting on AIs.''}
Such doubt is often grounded in the observation that comments under the sanme video tend to be similar:
\textit{``Everyone replies the same way -- it’s suspicious.''}
Some expressed their concerns with a sarcastic tone:
\textit{``First day being a bot, how did I do?''}

Further, comment sections often drift into unrelated or surreal topics, such as religious sermons, conspiracy remarks, or nonsensical humor. 
The collapse of topical focus further erodes the perception of coherence and authenticity:
\textit{``Not sure why folks talk about religion under an AI cat video.''}
\textit{``Imagine posting conspiracies under a cartoon.''}
\textit{``Why are people talking about politics here?''}
The absence of conversational coherence reinforces users’ suspicion that they inhabit an automated ecosystem.

\subsubsection{Ask for Likes, Nevertheless}

To maintain visibility, users imitate familiar social media behaviors -- asking for likes, tags, or reactions -- even when such features do not function on Sora, with the prevalence of bots. 
This performativity amplifies the perception of artificiality,
\textit{``Smash that heart button!''}
\textit{``All engagement here feels fake.''}
\textit{``Fake engagement just to look popular.''}
\textit{``Empty interaction everywhere.''}

\subsubsection{Watermarking as a Potential Solution to Authenticity}

Watermarking of AI-generated content was suggested as a potential solution to distinguishing AI-generated and real-life videos, and non-use of watermarking was questioned: \textit{``Why don’t these videos say sora on them?''}

\subsection{Contesting Platform Governance -- Negotiating Rules and Restrictions}

Sora discourages the use of real people's images as prompts, but some people nevertheless created videos in such a way, which was questioned by others, 
\textit{``This video is against the guidelines of this app.''}
\textit{``How can you create it?? Sora allow you?''} 

\subsubsection{Frustration Toward Over-regulation and Opaque Rules}

Some users express frustration toward the platform’s moderation system:  
\textit{``Remove all content violations, no more content violations!''} 
\textit{``Please fix your content policy, I tried to make an innocent video would not let me.''} 
These comments reveal the ongoing negotiation between user creativity and platform control.

Users also express irritation at unclear moderation criteria. 
Many accuse the system of unfairly blocking innocent or artistic content without explanations.
\textit{``Please fix your content policy -- I tried to make an innocent video.''}
\textit{``Got a violation for literally nothing.''}
\textit{``Why is this flagged?''}
\textit{``Sora needs transparent moderation.''}
Inconsistencies and frequent rule changes in the moderation system are another major complaint: 
\textit{``The rules keep changing.''}
\textit{``AI allowed this yesterday but not today.''}
The lack of appeal channels frustrates users,
\textit{``There’s no appeal process.''}
\textit{``Videos disappear with no explanation.''}
These frustrations reveal that many users do not reject moderation itself, but rather, they contest its opacity. 
Users want consistency, explanation, and control, not anarchy.

\subsubsection{Linguistic Evasion and Beyond}

While Sora’s moderation system attempts to limit inappropriate or copyrighted content, users treat these restrictions as puzzles to be solved rather than boundaries to obey. 
They share practical methods to evade detection, exchange strategies, and test the limits of what the system allows.
 
Many users learn that moderation operates through keyword filtering. 
They exploit this by deliberately misspelling, paraphrasing restricted terms, or using other linguistic approaches:
\begin{itemize}
    \item (Additional puctuations or spaces) \textit{``Use spaces between letters -- like F a m i l y G u y.''}
    \item (Typo) \textit{``When I do family guy I type ‘gamily fuy.’ Works every time.''}
    \item (Implicit expression) \textit{``Try saying ‘white female with orange hair’ instead of Lois Griffin.''}
    \item (Emoji) \textit{``Sometimes emojis help bypass filters.''}
    \item (Non-English Language) \textit{``Type in another language -- it confuses the system.''}
    \item (Avoid sensitive words) \textit{``Avoid words like ‘blood’—say ‘red water.’''}
    \item (Visual alternatives) \textit{``Just screenshot the text and upload it.''}
    \item (Rewording) \textit{``If you rewrite the sentence slightly, it passes.''}
\end{itemize}
This evolving lexicon of evasion functions shows that users collectively invent alternative vocabulary that adapts to the system’s filters.



\subsubsection{Peer Ethical Policing}

Interestingly, while users try to circumvent platform rules, they simultaneously enforce their own ethical boundaries through community norms.
\textit{``This ain’t how you respect the deceased.''}
\textit{``Stop using celebrity faces without consent.''}
\textit{``Be respectful -- it’s someone’s image.''}
\textit{``This is not okay, even if AI made it.''}
\textit{``It feels wrong to make people say things they never said.''}
\textit{``Respect privacy even in AI.'}'
These peer interventions demonstrate that community norms coexist with subversive tactics, and that users can both resist and self-regulate within the same space.

\section{Discussion}

Below, we discuss how users negotiate content authenticity, prompt ownership, and platform governance on/with Sora.
We propose design and policy implications to address the challenges in these negotiations, e.g., how platforms could serve as intermediaries that enable prompt monetization while preserving intellectual property.

\subsection{User Negotiation of Authenticity, Ownership, and Governance}

Across the literature, users increasingly appear as active agents who negotiate the socio-technical boundaries of generative AI systems \cite{wang2025excavating}. 
Rather than simply accepting platform-provided definitions of authenticity, users actively evaluate realism, cross-check sources, and share detection strategies \cite{farooq2025deciphering}. 
Our results echo such findings, with realism playing a crucial role in user negotiations of authenticity.
Users frequently evaluate microvisual details, such as lighting consistency, shadows, motion physics, and temporal coherence, to determine whether an AI-generated video ``could exist'' in the physical world. 
This form of everyday visual forensics illustrates that authenticity judgments are enacted through practical, user-driven evaluation practices rather than exclusively through expert verification.

Our findings show that embodied reactions are possible on viewing AI-generated video content and they illustrate that realism registers directly through the body, bypassing conscious reasoning. This appears to indicate that despite the back-and-forth conversation among users concerning its realism, Sora appears to have crossed the Uncanny Valley \cite{mori2012uncanny} for many users through its physiological impact.  

In our analysis, Sora users frame the platform as both a creative tool and an amplifier of deepfake risk. 
Users frequently debate whether hyper-realistic outputs should be celebrated as innovation or restricted due to their potential to spread misinformation and impersonate others.
As the distinction between real and synthetic, creator and algorithm, and user and bot continues to blur, Sora becomes a mirror of the broader digital condition, where authenticity is constantly negotiated, and trust must be redefined.
A practical design implication is for platforms to label AI-generated content and nudge users to distinguish between authentic and AI-generated videos, thereby improving users' AI literacy over time \cite{racine2025understanding}; however, such practices may conflict with corporate priorities, such as greater visibility for the videos or profits \cite{beyari2025role}.
Laws mandating the labeling of AI-generated content are plausible, especially as technology evolves and physical flaws in AI-generated videos are reduced.

In terms of ownership, users contest authorship through discourse on prompt originality, remix ethics, and creative labor. 
In the literature, psychological ownership mechanisms similarly explain why users feel entitled to credit and control even in algorithmically mediated creation environments \cite{wang2025excavating}. 
Legal ambiguity amplifies these authorship negotiations, as formal copyright protection often fails to reflect lived user experiences \cite{wang2025research, khadka2025navigating}.
The tension between formal copyright regimes and informal user norms illustrates how ownership is socially negotiated in generative platforms. 
In our analysis, users assert moral and creative rights over their contributions, and publicly debate who truly ``owns'' an AI-generated video: the prompt writer, or the platform itself.
As content creators create more sophisticated prompts for generating videos, the debate over prompt ownership may receive even greater attention in the future.  

The rapid diffusion of generative AI has prompted significant interest in regulatory and governance frameworks. 
Mansouri et al. propose a unified ethical and legal framework for AI accountability, emphasizing explainability, transparency, and harm mitigation across sectors \cite{mansouri2025ethical}. 
Abishanth and Banerjee similarly highlight emerging governance challenges, noting that regulatory responses often lag behind technological innovation \cite{abishanth2025study}.

At the institutional level, the European Union AI Act is among the most comprehensive attempts to regulate high-risk AI applications \cite{busch2024navigating}. 
While the Act primarily targets domains such as healthcare and public safety, its risk-based approach has implications for generative video systems that may be used for manipulation, surveillance, or biometric deception.

Auditing has also emerged as a key governance mechanism. 
Li and Goel demonstrate that AI auditability remains uneven across platforms, with limited access to training data, model architecture, and content moderation processes \cite{li2025making}. 
This opacity complicates public trust and weakens external oversight. 
Ortega-Bolaños et al. further show that although numerous ethical assessment tools exist, their practical integration into platform governance remains inconsistent \cite{ortega2024applying}.

In our results, we found that governance is negotiated through everyday interaction. 
Users not only comply with or resist moderation policies but also actively shape platform norms by discouraging harmful uses, reporting misuse, and sharing ethical expectations within community spaces. 
These micro-level governance practices highlight how regulatory power is distributed across platforms, legal institutions, and user collectives.
Future research should explore the development of participatory governance frameworks that allow users to directly influence platform policy through structured feedback, community moderation, and co-regulatory design.

\subsection{Keeping Prompts Private: Authorship on Sora}

On Sora, prompts have become a form of creative property, which helps explain users’ reluctance to share them. 
A well-crafted prompt can generate viral, monetizable videos; for many, it represents accumulated craft and intuition rather than casual text. 
Sharing a prompt would mean giving away the creative logic that defines one’s style. 
This perception echoes emerging marketplaces such as PromptBase \cite{promptbase2025}, where creators sell prompt templates and profit from their design. 
Within this new economy, prompts carry both symbolic and economic value, and withholding them protects intellectual ownership and competitive advantage. 

The emergence of prompts as intellectual property has been discussed in prior literature on text-based large language models~\cite{yao2024promptcare, reissinger2025safer}, arguing that prompts themselves warrant safeguards against unauthorized reuse \cite{van2023protect}.
However, the creative workflow in text-based AI tools fundamentally differs from that in video generation platforms like Sora.
When using LLMs for writing or coding, users typically employ a hybrid workflow: the model produces an initial draft, which the human then substantially revises, restructures, and refines.
The final artifact thus reflects a collaborative synthesis of algorithmic suggestion and human editorial judgment.
In contrast, video generation imposes a fundamentally different constraint.
Unlike text, which can be edited at the word or sentence level, AI-generated video cannot be manipulated with the same granularity using conventional tools: even professional video editing software cannot meaningfully alter the underlying content of a generated scene.
This asymmetry shifts the locus of creative contribution almost entirely to the prompt.

This structural limitation elevates the intellectual property significance of prompts in video generation contexts.
As our findings illustrate, users on Sora treat prompts as ``creative capital'' and actively protect, request, and even commodify them (e.g., ``Protect your prompt, people will steal it, '' ``You should sell that prompt'').
Unlike text-based workflows, where the value proposition is distributed across model output and human refinement, Sora's creative pipeline collapses this distribution into a singular dependency on prompt engineering.
Consequently, users who seek to monetize their content or gain visibility on the platform must rely almost exclusively on the quality and originality of their prompts.
In other words, there is no downstream editing stage to differentiate one's work.
This amplifies concerns around prompt plagiarism and attribution, as evidenced by accusations such as ``That's my exact prompt!'' and ``Wow thanks for stealing my content.''
We argue that while prompt-as-IP has been acknowledged in prior work \cite{van2023protect}, the inherent uneditable nature of video outputs makes this concern substantially more acute in generative video platforms, warranting dedicated legal and platform-level attention.

Users also voice privacy and control concerns when negotiating authorship with AI. 
Prompts entered into AI systems may be stored or reused for model training, raising uncertainty about how creative input is appropriated \cite{privacy_prompt_data, uiuc_genai_privacy}. 
These behaviors resonate with Dang et al., who found that composing explicit, non-diegetic prompts can interrupt creative rhythm and demand substantial cognitive effort, turning prompt writing into a form of authorship \cite{dang2023choice}. 
On Sora, that authorship becomes a marker of expertise, distinguishing experienced creators from casual participants. 
Refusing to share prompts or authorship with AI is therefore more than secrecy or competition -- it is boundary work through which users manage authorship, value, and trust in an ecosystem increasingly shaped by algorithmic production.

One potential design implication emerges from Sora's existing remix functionality.
Rather than requiring creators to directly disclose or sell their prompts, which risks unauthorized duplication, the platform could serve as an intermediary that enables prompt monetization while preserving intellectual property.
Specifically, by allowing other users to build on or remix existing videos without exposing the underlying prompt, Sora could create an economic model in which original creators receive attribution or compensation whenever their work serves as a foundation for derivative content.
This approach mirrors licensing structures in traditional creative industries, where derivative works generate royalties for original authors \cite{morrison2006derivers}.
Such a mechanism would address the community's concerns about prompt theft (e.g., ``Tag me if you remix, don't pretend it's yours'') while still fostering the collaborative culture that platforms like Sora seek to cultivate.
Future research should explore how platform affordances can balance open creativity with intellectual property protection, and how new interaction paradigms, such as iterative prompt refinement or template libraries, can scaffold more effective human-AI creative partnerships in generative video contexts.
Legal scholarship should advance enforceable hybrid authorship models that account for shared creative agency between humans and AI systems. 

\subsection{Limitations and Future Work}

This study has several limitations that open opportunities for future research. 
First, our analysis is based solely on publicly visible user comments under Sora-generated videos. 
While such comments provide valuable insight into how users collectively evaluate realism, negotiate authorship, and discuss governance issues, they do not capture private reactions, silent viewers, or users who engage only through viewing rather than commenting. 
Future work could incorporate interviews or surveys to better understand the broader population of users who may hold different perceptions but do not express them publicly.

Second, our dataset focuses on 41 videos with at least 10 comments, which allowed us to examine sufficiently rich discussions but may not reflect the full diversity of content or user experiences on the platform. 
As Sora becomes more widely adopted, future research should examine larger, more heterogeneous datasets, such as less-commented videos.

Third, our findings are shaped by Sora's early-stage nature as a platform. 
Users’ perceptions of realism, creativity, and governance may shift as the tool matures, as moderation policies stabilize, and as commercial use cases expand. 
Longitudinal studies could investigate how community norms, prompting practices, and governance expectations evolve over time.
Digital ethnography, platform scraping, and experimental exposure studies would be particularly valuable for capturing these dynamic processes.

Fourth, this study focuses on English-language comments, which limits our ability to capture perspectives from international communities. 
Given that users in our dataset already employed multilingual prompting techniques to circumvent restrictions, future work could conduct cross-cultural or multilingual analyses to examine how norms differ across cultures.

\section{Conclusion}

This study shows that users on Sora actively negotiate authenticity, authorship, and governance as AI-generated video becomes increasingly realistic. 
Through qualitative analysis of user comments, we found that people verify realism through physical details, treat prompts as creative property, and contest moderation by challenging opaque rules while simultaneously enforcing community-driven ethical expectations. 
These findings highlight that governance in generative video ecosystems is co-constructed by platforms and users. 
Future work should examine broader and more diverse user groups to understand how these norms evolve as AI video technologies mature.

\section*{Generative AI Use}

During the preparation of this work, the authors used ChatGPT for writing assistance, such as brainstorming high-level insights from the data and proofreading. 
After using this tool, the authors reviewed and edited the content as needed and take full responsibility for the publication's content.


\bibliographystyle{IEEEtran}
\bibliography{references}

\end{document}